\documentclass[aps,
pre,
twocolumn,
longbibliography,
superscriptaddress]{revtex4-1}

\usepackage{graphicx}
\usepackage{subfigure}
\usepackage{amssymb,amsmath,physics}
\usepackage{float}
\usepackage{xcolor}
\allowdisplaybreaks

\begin{document}

\renewcommand{\floatpagefraction}{0.9}

\title{Integrated Lennard-Jones Potential between a Sphere and a Thin Rod}

\author{Junwen Wang}
\affiliation{Department of Mechanical Engineering, Virginia Tech, Blacksburg, Virginia 24061, USA}
\affiliation{Center for Soft Matter and Biological Physics, Virginia Tech, Blacksburg, Virginia 24061, USA}
\affiliation{Macromolecules Innovation Institute, Virginia Tech, Blacksburg, Virginia 24061, USA}
\author{Shengfeng Cheng}
\email{chengsf@vt.edu}
\affiliation{Department of Physics, Virginia Tech, Blacksburg, Virginia 24061, USA}
\affiliation{Center for Soft Matter and Biological Physics, Virginia Tech, Blacksburg, Virginia 24061, USA}
\affiliation{Macromolecules Innovation Institute, Virginia Tech, Blacksburg, Virginia 24061, USA}
\affiliation{Department of Mechanical Engineering, Virginia Tech, Blacksburg, Virginia 24061, USA}

\begin{abstract}
A compact analytical form is derived via an integration approach for the interaction between a sphere and a thin rod of both finite and infinite lengths, with each object treated as a continuous medium of materials points interacting by the Lennard-Jones 12-6 potential. Expressions for the resultant forces and torques are obtained. Various asymptotic limits of the analytical sphere-rod potential are discussed.
\end{abstract}

\date{\today}


\maketitle

\section{Introduction}

The Lennard-Jones (LJ) 12-6 potential is one of the most frequently used functional forms to represent interatomic and intermolecular interactions.\cite{Fischer2023, Allen2017book} As a natural extension, integrated LJ potentials between condensed bodies with various geometrical shapes can find applications in a wide range of computational studies and theoretical analyses because of their simplicity,\cite{Everaers2003, Vesely2006} though pairwise additivity is assumed in the so-called Hamaker approach.\cite{Hamaker1937} Furthermore, the parameters (e.g., the Hamaker constant setting the interaction strength) in an integrated potential can be tuned to match realistic cases, enhancing the applicability of such potentials. Integrated forms of the LJ 12-6 potential have been derived for a few spherical and planar geometries.\cite{deRocco1960, Abraham1977, Abraham1978, Magda1985, Everaers2003} The integrated potential between two spheres has been implemented in the Large-scale Atomic/Molecular Massively Parallel Simulator (LAMMPS).\cite{LAMMPS_COLLOID} The LJ potential has also been integrated between a point particle and a plane/half-space and implemented in LAMMPS as various wall potentials.\cite{LAMMPS_WALL}

In addition to spheres, cylindrical objects are abundant in synthetic and natural systems,\cite{Israelachvili2011} including liquid crystal molecules, colloidal nanorods, nanopillars, carbon nanotubes, nanowires, biofilaments (e.g., microtubules), and rod-shaped virus (e.g., tobacco mosaic virus) and microorganisms (e.g., \textit{Escherichia coli} bacteria). In general, it is more challenging to integrate the LJ potential for cylinders. Most attempts have been made in studies of carbon nanotubes \cite{Henrard1999, Girifalco2000, Sun2006PRB, Popescu2008, Lu2009APL, Zhbanov2010, Pogorelov2012} and some are limited to the van der Waals (vdW) attraction only. The full LJ 12-6 potential was integrated by de Rocco and Hoover for two thin rods in either collinear or parallel configurations.\cite{deRocco1960} Hamady et al. derived an analytical expression for the interaction between a nanorod and a three-dimensional half-space filled with LJ point particles.\cite{Hamady2013} An approximate form of the integrated LJ potential was proposed by Vesely for two sticks in more general settings.\cite{Vesely2006} Recently, full analytical forms for both $1/r^6$ attraction and $1/r^{12}$ repulsion have been obtained by Wang et al. for thin LJ rods in arbitrary 3-dimensional arrangements,\cite{Wang2024} by means of Ostrogradski's integration method.\cite{Ostrogradski1845a,Ostrogradski1845b}

To study mixtures of spheres and rods theoretically and computationally, it is necessary to include their mutual interactions. Several results have been previously reported in the literature on the rod-sphere interactions. Rosenfeld and Wasan obtained the exact result on the nonretarded vdW attraction between a sphere and an infinitely long cylinder with a finite radius by integrating the $1/r^6$ potential.\cite{Rosenfeld1974} Kirsch later confirmed this result and further obtained the compact expression for the retarded case (i.e., the integrated form of the $1/r^7$ attractive potential).\cite{Kirsch2003} Gu and Li studied both retarded and nonretarded vdW interaction between a sphere and a cylinder with a finite length and cross-section by combining analytical and numerical integrations.\cite{Gu1999} Montgomery et al. calculated the dispersion forces for several nontraditional geometries, including the case of a sphere and an infinite cylinder.\cite{Montgomery2000} He et al. obtained an analytical expression for the vdW attraction between a nanoparticle and a nanorod with a finite length and radius under certain approximations.\cite{He2012PCCP} However, a compact form of the integrated LJ potential between a sphere and a rod has been elusive.

Here we report a fully analytical form of the interaction between a sphere and a thin rod with either finite or infinite lengths in an arbitrary configuration by integrating the LJ 12-6 potential between a pair of particle. The sphere is treated as a continuum and the thin rod is modeled as a material line consisting of LJ particles. The integrated sphere-rod potential is expressed as an analytical function form and the associated expressions for forces and torques are also presented. These forms can be used in theoretical analysis and computational modeling of sphere-rod mixtures.

\section{Theoretical Model of Sphere-Rod Interactions}

\subsection{Integrated Sphere-Point Potential}

The LJ 12-6 potential between two point particles reads 
\begin{equation}
    \label{eq:lj_potential}
    U_\text{LJ}(r) = 4\epsilon \left[\left( \frac{\sigma}{r}\right)^{12} - \left(\frac{\sigma}{r}\right)^6 \right]~,
\end{equation}
where $\epsilon$ is an energy scale, $\sigma$ is a length scale, and $r$ is the distance between particles. The LJ 12-6 potential has been integrated between a point particle and a sphere regarded as a uniform distribution of LJ particles at a number density of $1.0\sigma^{-3}$.\cite{Everaers2003} The resulting sphere-point potential can be expressed as
\begin{eqnarray}
    \label{eq:sphere_point_potential}
    U_\text{SP}(r) &=& \frac{2a^3\sigma^3 A_{cs}}{9} \left[ \frac{\left( 5 a^6 +45 a^4 r^2 + 63 a^2 r^4 +15 r^6\right)\sigma^6}{15 \left(r^2-a^2\right)^9} \right.\nonumber \\
    & & \left. - \frac{1}{\left(r^2-a^2\right)^3}\right]~,
\end{eqnarray}
where $r$ is the center-to-center distance between the point particle and sphere, $a$ is the sphere radius, and $A_{cs} = 24\pi\epsilon$ is a Hamaker constant setting the interaction strength. Clearly, $r>a$ is required in Eq.~(\ref{eq:sphere_point_potential}) as the point particle cannot overlap with the sphere.

\subsection{Integrated Sphere-Rod Potential}

\begin{figure}[H]
    \centering
    \includegraphics[width=0.45\textwidth]{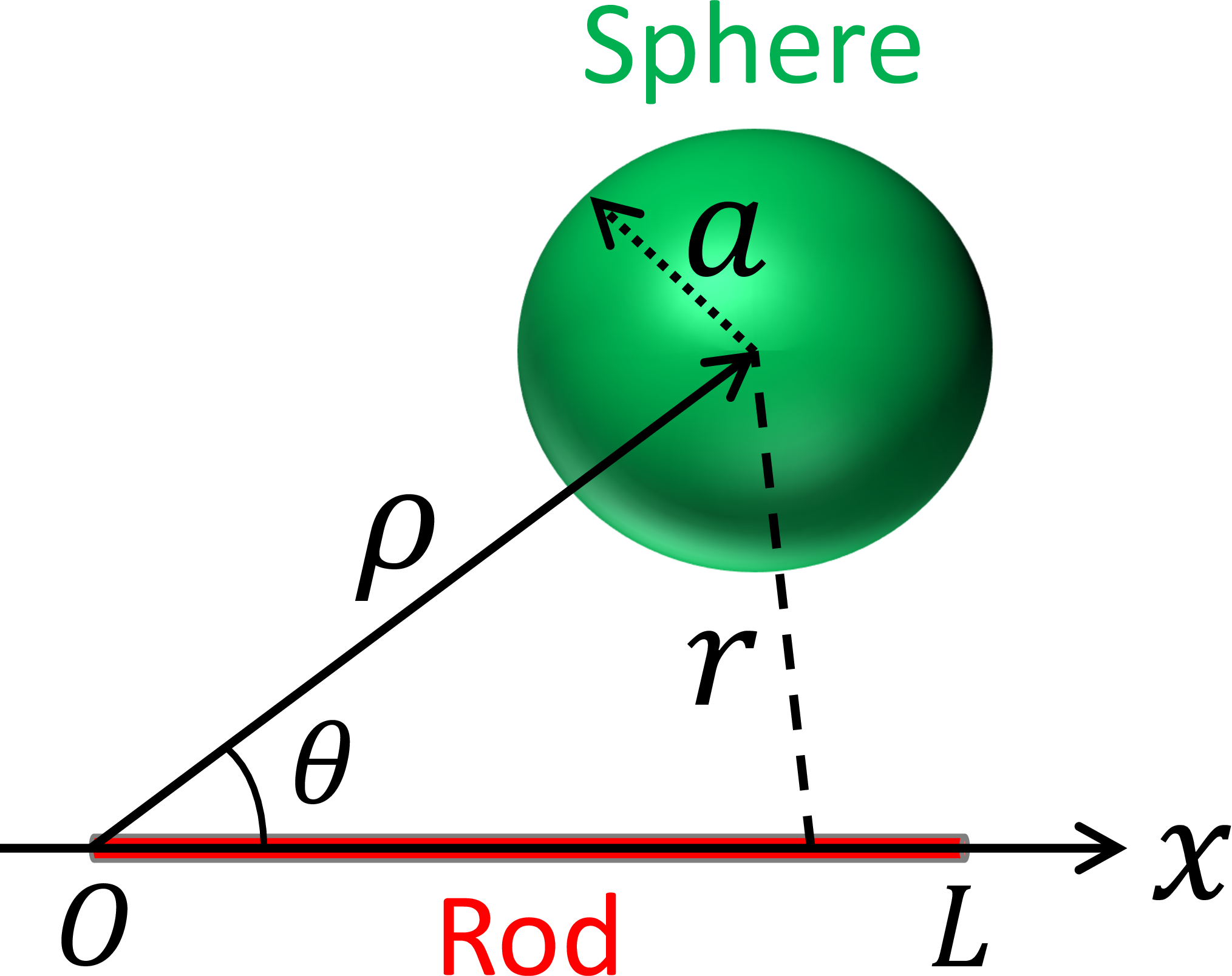}
    \caption{A polar coordinate system describing a general configuration of a sphere and a thin rod.}
    \label{fg:rod-sphere-geo}
\end{figure}

The sphere-point potential can be integrated further to obtain the interaction between a sphere and a thin rod. A general sphere-rod configuration is shown in Fig.~\ref{fg:rod-sphere-geo}. By setting the $x$-axis along the central axis of the rod and choosing one end of the rod as the origin, we can build a polar coordinate system with the center of the sphere located at $(\rho,\theta)$. It is always possible to build such a frame with $0 \le \theta \le \pi$. The interaction potential between the sphere and rod can then be denoted as a function $W(\rho,\theta)$, which represents the integrated form of the sphere-point interaction potential in Eq.~(\ref{eq:sphere_point_potential}).
\begin{eqnarray}
    \label{eq:int_potential_one}
    W(\rho,\theta) &=& \lambda \int_{0}^{L} U_\text{SP}(r)\vert_{r=\sqrt{x^2 + \rho^2 -2 x \rho \cos\theta}}~dx~,
\end{eqnarray}
where $L$ is the length of the rod, $\lambda$ the line number density of LJ material points that the rod consists of.

With a change of variables, $y=x-\rho\cos\theta$ and $h = \rho\sin\theta$, the integral in Eq.~(\ref{eq:int_potential_one}) can be transformed into
\begin{eqnarray}
    \label{eq:int_potential_two}
    W(\rho,\theta) &=& \lambda \int_{0}^{L} U_\text{SP}(r)\vert_{r=\sqrt{y^2 + h^2}}~dx~.
\end{eqnarray}
Considering the form of $U_\text{SP}(r)$ in Eq.~(\ref{eq:sphere_point_potential}), this integral is an integral of rational functions, which can be evaluated using Ostrogradski's method.\cite{Ostrogradski1845a, Ostrogradski1845b} The result can be written as
\begin{equation}
    \label{eq:int_pot_general}
    W(\rho,\theta) = \frac{2\lambda a^3\sigma^3 A_{cs}}{9} \left[ G(L,\rho,\theta) - G(0,\rho,\theta)\right]
\end{equation}
where the function $G(x,\rho,\theta)$ is given by
\begin{widetext}
\begin{eqnarray}
    \label{eq:int_sphere_rod_pot}
    G(x,\rho,\theta) &=& \sigma^6 \left[ \frac{8 a^6 y}{15 \left(h^2-a^2\right) \left(h^2+y^2-a^2\right)^8}  -\frac{4 y \left(4 a^6-9 a^4 h^2\right)}{35 \left(h^2-a^2\right)^2 \left(h^2+y^2-a^2\right)^7}  +\frac{y \left(11 a^6-9 a^4 h^2+63 a^2 h^4\right)}{105 \left(h^2-a^2\right)^3 \left(h^2+y^2-a^2\right)^6} \right. \nonumber \\
    & & + y \left(16 a^6+216 a^4 h^2+378 a^2 h^4+105 h^6\right) \times \left( \frac{1 }{1050 \left(h^2-a^2\right)^4 \left(h^2+y^2-a^2\right)^5} \right. \nonumber \\
    & & +\frac{3 }{2800 \left(h^2-a^2\right)^5 \left(h^2+y^2-a^2\right)^4} + \frac{1 }{800 \left(h^2-a^2\right)^6 \left(h^2+y^2-a^2\right)^3} \nonumber \\
    & & \left. +\frac{1 }{640 \left(h^2-a^2\right)^7 \left(h^2+y^2-a^2\right)^2} + \frac{3 }{1280 \left(h^2-a^2\right)^8 \left(h^2+y^2-a^2\right)} \right) \nonumber \\
    & & \left. + \frac{3 \left(16 a^6+216 a^4 h^2+378 a^2 h^4+105 h^6\right)}{1280 \left(h^2-a^2\right)^{17/2}} \arctan\left(\frac{y}{\sqrt{h^2-a^2}}\right) \right] \nonumber \\
    & &  - \left[ \frac{y}{4 \left(h^2-a^2\right) \left(h^2+y^2-a^2\right)^2} + \frac{3 y}{8 \left(h^2-a^2\right)^2 \left(h^2+y^2-a^2\right)} \right. \nonumber \\
    & & \left. + \frac{3}{8 \left(h^2-a^2\right)^{5/2}} \arctan \left(\frac{y}{\sqrt{h^2-a^2}}\right) \right]~,
\end{eqnarray}
\end{widetext}
where $y=x-\rho\cos\theta$ and $h=\rho\sin\theta$. Here the repulsive and attractive components of the interaction are clearly separated.

Equation (\ref{eq:int_sphere_rod_pot}) holds as long as $h>a$, i.e., the $x$-axis in Fig.~\ref{fg:rod-sphere-geo} is outside the sphere. In the case of $h<a$, the sphere intersects with the $x$-axis (i.e., the line along the central axis of the rod). The argument of the arctangent function in Eq.~(\ref{eq:int_sphere_rod_pot}), $y/\sqrt{h^2-a^2}$ becomes a complex number. This is expected as the integrated sphere-rod potential should diverge when the two overlap. However, the potential should still be finite as long as the rod is outside the sphere. Therefore, for $h<a$, the arctangen term need to be transformed into
\begin{eqnarray}
    \label{eq:arctangent_transform}
    & & \frac{1}{\sqrt{h^2-a^2}}\arctan \left(\frac{y}{\sqrt{h^2-a^2}}\right) \nonumber \\
    &= & \frac{1}{2\sqrt{a^2-h^2}} \ln \left( \frac{\sqrt{a^2-h^2}-y}{\sqrt{a^2-h^2}+y}\right)~.
\end{eqnarray}
It is easy to show that for a rod outside the sphere with $h<a$, the argument of the logarithmic function in Eq.~(\ref{eq:arctangent_transform}) is negative for $-\rho\cos\theta \le y \le L - \rho\cos\theta $. Therefore, directly using Eq.~(\ref{eq:arctangent_transform}) still yields a complex value for $G(x,\rho,\theta)$. However, since only $G(L,\rho,\theta) - G(0,\rho,\theta)$ enters the integrated potential, the relative term involves the following expression
\begin{eqnarray}
    \label{eq:log_transform}
    & & \ln \left( \frac{\sqrt{a^2-\rho^2\sin^2\theta}-L+\rho\cos\theta}{\sqrt{a^2-\rho^2\sin^2\theta}+L-\rho\cos\theta}\right) \nonumber \\
    & & - \ln \left( \frac{\sqrt{a^2-\rho^2\sin^2\theta}+\rho\cos\theta}{\sqrt{a^2-\rho^2\sin^2\theta}-\rho\cos\theta}\right) \nonumber \\
    &=& \ln \left( \frac{a^2-\rho^2-L\sqrt{a^2-\rho^2\sin^2\theta} + L\rho\cos\theta}{a^2-\rho^2+L\sqrt{a^2-\rho^2\sin^2\theta} + L\rho\cos\theta} \right)~.
\end{eqnarray}
The argument of the logarithmic function in the last line of Eq.~(\ref{eq:log_transform}) is positive and the function value is thus real.

To make it clear, in the case of $h<a$, the integrated potential can be explicitly written as
\begin{equation}
    \label{eq:int_pot_general_two}
    W(\rho,\theta) = \frac{2\lambda a^3\sigma^3 A_{cs}}{9} \left[ Q(L,\rho,\theta) - Q(0,\rho,\theta) + P(\rho, \theta)\right]~,
\end{equation}
with
\begin{widetext}
\begin{eqnarray}
    \label{eq:int_sphere_rod_pot_two}
    Q(x,\rho,\theta) &=& \sigma^6 \left[ \frac{8 a^6 y}{15 \left(h^2-a^2\right) \left(h^2+y^2-a^2\right)^8}  -\frac{4 y \left(4 a^6-9 a^4 h^2\right)}{35 \left(h^2-a^2\right)^2 \left(h^2+y^2-a^2\right)^7}  +\frac{y \left(11 a^6-9 a^4 h^2+63 a^2 h^4\right)}{105 \left(h^2-a^2\right)^3 \left(h^2+y^2-a^2\right)^6} \right. \nonumber \\
    & & + y \left(16 a^6+216 a^4 h^2+378 a^2 h^4+105 h^6\right) \times \left( \frac{1 }{1050 \left(h^2-a^2\right)^4 \left(h^2+y^2-a^2\right)^5} \right. \nonumber \\
    & & +\frac{3 }{2800 \left(h^2-a^2\right)^5 \left(h^2+y^2-a^2\right)^4} + \frac{1 }{800 \left(h^2-a^2\right)^6 \left(h^2+y^2-a^2\right)^3} \nonumber \\
    & & \left. \left. +\frac{1 }{640 \left(h^2-a^2\right)^7 \left(h^2+y^2-a^2\right)^2} + \frac{3 }{1280 \left(h^2-a^2\right)^8 \left(h^2+y^2-a^2\right)} \right)  \right] \nonumber \\
    & &  - \left[ \frac{y}{4 \left(h^2-a^2\right) \left(h^2+y^2-a^2\right)^2} + \frac{3 y}{8 \left(h^2-a^2\right)^2 \left(h^2+y^2-a^2\right)} \right]~,
\end{eqnarray}
\end{widetext}
and
\begin{eqnarray}
    \label{eq:int_sphere_rod_pot_two_extra}
    P(\rho,\theta) &=& \left[ \frac{3\sigma^6 \left(16 a^6+216 a^4 h^2+378 a^2 h^4+105 h^6 \right)}{2560 \left(a^2- h^2\right)^{17/2}} \right. \nonumber \\
    & & \left. - \frac{3}{16 \left(a^2 - h^2\right)^{5/2}} \right] \times \nonumber \\
    & & \ln \left( \frac{a^2-\rho^2-L\sqrt{a^2-h^2} + L\rho\cos\theta}{a^2-\rho^2+L\sqrt{a^2-h^2} + L\rho\cos\theta} \right)~,
\end{eqnarray}
where $y=x-\rho\cos\theta$ and $h=\rho\sin\theta$.

In the case of $h=a$, i.e., when the $x$-axis in Fig.~\ref{fg:rod-sphere-geo} is tangent to the sphere, the integrated potential between the sphere and rod can be easily evaluated to be
\begin{eqnarray}
    \label{eq:int_sphere_rod_tangent}
    W(\rho,\theta) &=& \frac{2\lambda a^3\sigma^3 A_{cs}}{9} \left[
    \frac{1}{5y^5}-\frac{\sigma^6}{15} \left( \frac{128a^6}{17y^{17}} + \frac{72a^4}{5y^{15}} \right. \right. \nonumber \\
    & & \left. \left. + \frac{108a^2}{13y^{13}}+ \frac{15}{11y^{11}}\right) \right]^{y= L-\rho\cos\theta}_{y= -\rho\cos\theta}~.
\end{eqnarray}
It can be shown that Equation (\ref{eq:int_sphere_rod_tangent}) is the asymptotic form of Eq.~(\ref{eq:int_sphere_rod_pot}) in the limit of $h=a$, except for a singular term that only depends on $h$ and is thus canceled out during the subtraction process in Eqs.~(\ref{eq:int_pot_general}) or (\ref{eq:int_pot_general_two}).

For an infinitely long thin rod, the integrated sphere-rod potential can be written as
\begin{eqnarray}
    \label{eq:int_pot_inf_rod}
    W(\rho,\theta) &=& \frac{2\lambda a^3\sigma^3 A_{cs}}{9} \lim_{L\rightarrow \infty}\left[ G(L,\rho,\theta) - G(0,\rho,\theta) \right. \nonumber \\
    & & \left. + G(L,\rho, \pi-\theta) - G(0,\rho,\pi-\theta) \right]~.
\end{eqnarray}
Note that $G(0,\rho,\theta)=-G(0,\rho,\pi-\theta)$ and
\begin{eqnarray}
    & & \lim_{L\rightarrow \infty} G(L,\rho,\theta) \nonumber \\
    &=& \lim_{L\rightarrow \infty} G(L,\rho,\pi-\theta) \nonumber \\
    &=& \frac{3 \pi \sigma^6 \left(16 a^6+216 a^4 h^2+378 a^2 h^4+105 h^6\right)}{2560 \left(h^2-a^2\right)^{17/2}} \nonumber \\
    & & - \frac{3\pi}{16 \left(h^2-a^2\right)^{5/2}}~.
\end{eqnarray}
The integrated LJ potential between a sphere of radius $a$ and an infinite thin rod thus reads
\begin{widetext}
\begin{equation}
    W(h) = \frac{\pi \lambda a^3\sigma^3 A_{cs}}{3} \left[ \frac{ \sigma^6 \left(16 a^6+216 a^4 h^2+378 a^2 h^4+105 h^6\right)}{640 \left(h^2-a^2\right)^{17/2}} - \frac{1}{4 \left(h^2-a^2\right)^{5/2}} \right]~.
\end{equation}
\end{widetext}
As expected, the potential in this case only depends on $h$, the distance between the sphere center and the central axis of the rod. Obviously, in this case $h>a$ is required. The force between the sphere and rod can be easily computed from $F(h) = -\text{d}W(h)/\text{d}h$. For example, the attractive component is
\begin{equation}
    \frac{5\pi \lambda a^3\sigma^3 A_{cs}}{12} \frac{h}{\left(h^2-a^2\right)^{7/2}}~,
\end{equation}
which is identical to the thin-rod limit of the result obtained by Rosenfeld and Wasan.\cite{Rosenfeld1974}

\begin{figure*}[htb]
    \centering
    \includegraphics[width=1\textwidth]{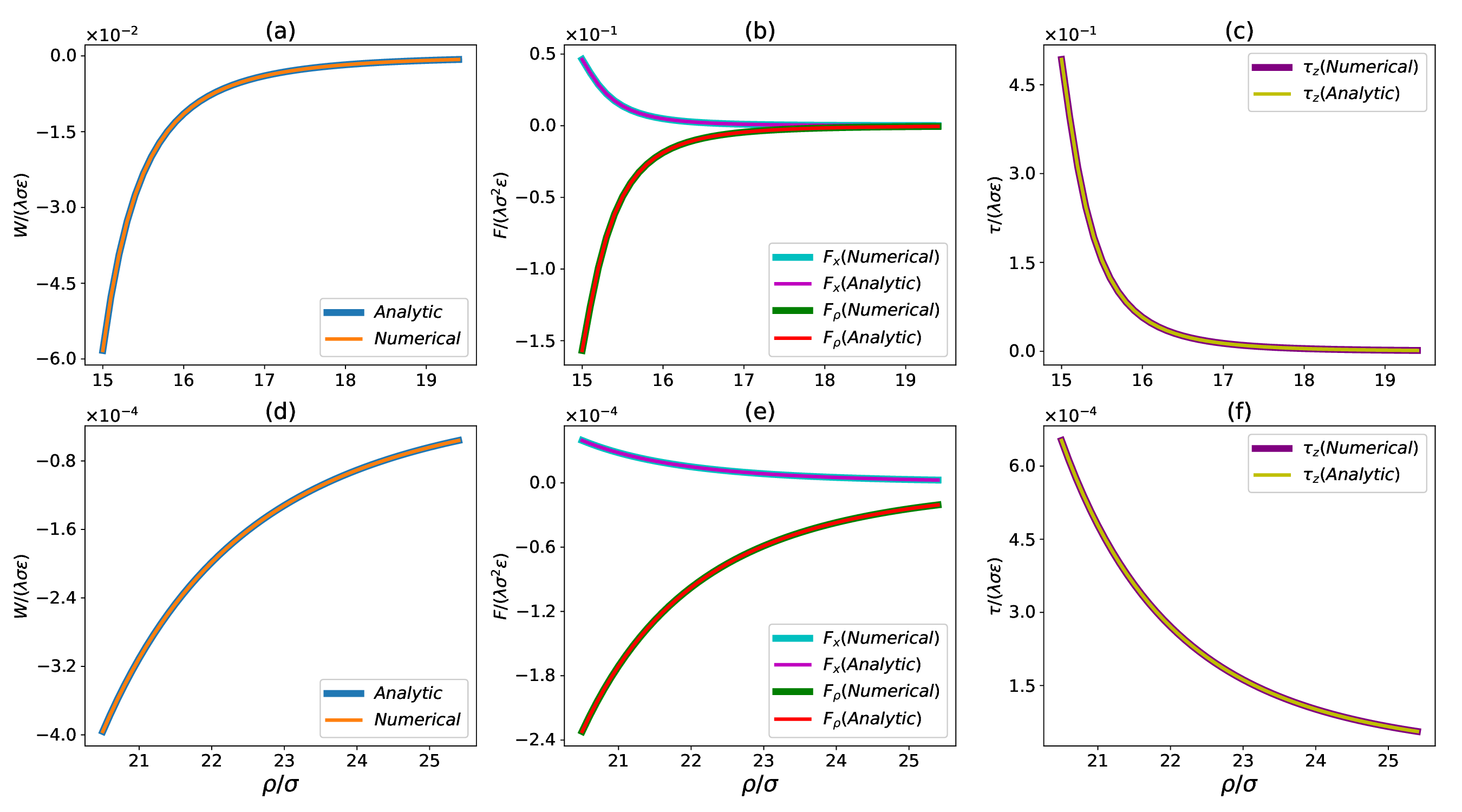}
    \caption{Comparison of analytical and numerical results on the integrated potentials ($W$) [(a) and (d)], force components ($F_x$ and $F_\rho$) [(b) and (e)], and torque on the rod ($\tau$) [(c) and (f)] vs. $\rho$. The results are for a sphere with $a = 10\sigma$ and a thin rod with $L=5\sigma$ at $\theta = \pi/6$. The top row is for the case with $\rho\sin\theta<a$ while the bottom row is for $\rho\sin\theta>a$.}
    \label{fg:SphereRod_comparison}
\end{figure*}

\subsection{Forces and Torques in Sphere-Rod Interactions}

Using the coordinate system defined in Fig.~\ref{fg:rod-sphere-geo}, the force on the sphere can be computed from the integrated potential, $W(\rho,\theta)$, as
\begin{equation}
    \vb{F}_\text{S} = \frac{1}{\rho \sin\theta}\frac{\partial W}{ \partial \theta} \vb{n}_x - \left( \frac{\partial W}{\partial \rho} + \frac{\cos \theta}{\rho \sin\theta}\frac{\partial W}{ \partial \theta} \right) \vb{n}_\rho~.
\end{equation}
The force on the rod is $\vb{F}_\text{R} = -\vb{F}_\text{S}$ from the Newton's third law.

The torque on the rod is
\begin{equation}
    \boldsymbol{\tau}_\text{Rz} = \left[ -\frac{L}{2} \sin\theta \frac{\partial W}{\partial \rho} + \left( 1- \frac{L}{2\rho} \cos\theta \right) \frac{\partial W}{\partial \theta} \right]\vb{n}_z~,
\end{equation}
where $\vb{n}_z \equiv \frac{1}{\sin\theta} \vb{n}_x\times \vb{n}_\rho$. The torque is 0 for $\theta =0$ and $\pi$ where the central axis of the rod passes through the sphere center. Furthermore, when $\rho = L/(2\cos\theta)$, i.e., when the sphere center is on the bisector of the rod, the torque on the rod is also 0. These limits are encoded in the symmetry of the sphere-rod system considered here.

\section{Verification of Analytical Results}

The analytical results on the integrated potentials, forces, and torques, which involve lengthy but compact expressions, are compared to the results from numerically integrating the LJ 12-6 potential for a sphere-rod system. A perfect agreement is found in all cases. Some examples are shown in Fig.~\ref{fg:SphereRod_comparison}. The comparison further validates our strategy [Eqs.~(\ref{eq:int_pot_general_two}), (\ref{eq:int_sphere_rod_pot_two}), and (\ref{eq:int_sphere_rod_pot_two_extra})] of dealing with the special situations where $\rho\sin\theta < a$, i.e., where the central axis of the rod intersects with the sphere. In these situations, the integrated potential is still real and finite as long as the rod is outside the sphere. The results in Fig.~\ref{fg:SphereRod_comparison} confirms that this is indeed the case if the operations in Eqs.~(\ref{eq:arctangent_transform}) and (\ref{eq:log_transform}) are adopted.

\section{Conclusions}

The Lennard-Jones (LJ) 12-6 potential has been successfully integrated for a sphere and a thin rod of both finite and infinite lengths in arbitrary 3-dimensional configurations, with the sphere and rod modeled as continuous media of LJ material points. The result has been expressed in a compact analytical form. The expressions of forces and torques are also presented. The integrated sphere-rod potential can be used for theoretical descriptions and computational modeling of soft matter systems involving mixtures of cylindrical and spherical objects.

\section*{Acknowledgements}
This material is based upon work supported by the National Science Foundation under Grant No. DMR-1944887. The authors acknowledge Advanced Research Computing at Virginia Tech (URL: http://www.arc.vt.edu) for providing computational resources and technical support that have contributed to the results reported within this paper.


%

\end{document}